\documentclass{article}

\usepackage{epsf}      
\usepackage{feynmp}    
\usepackage{amssymb}   
\usepackage{cite}

\skip\footins 39pt plus 4pt minus 2pt
\def\footnoterule{\kern-19pt\hrule width.5in\kern18.6pt}

\newcommand{\eq}[1]{equation~(\ref{#1})}
\newcommand{\Eq}[1]{Equation~(\ref{#1})}
\newcommand{\eqs}[2]{equations~(\ref{#1}) and~(\ref{#2})}

\newcommand{\fig}[1]{Figure~\ref{#1}}

\newcommand{\GLam}{G_{\Lambda}}
\newcommand{\GLamh}{G_{{\Lambda h}}}

\newcommand{\be}{\begin{equation}}
\newcommand{\ee}{\end{equation}}

\newcommand{\bea}{\begin{eqnarray}}
\newcommand{\eea}{\end{eqnarray}}

\begin{document}

\renewcommand{\topfraction}{0.99}
\renewcommand{\bottomfraction}{0.99}
\twocolumn[\hsize\textwidth\columnwidth\hsize\csname 
@twocolumnfalse\endcsname

\begin{center}
{\Large \bf The Sinc Function Representation and Three-Loop Master
Diagrams} \\
\bigskip
{\large Richard Easther, Gerald Guralnik, and Stephen Hahn} \\
\smallskip
{\small Department of Physics,  Brown University, 
Providence, RI  02912, USA.}\\
\medskip

\end{center}
\begin{abstract}
  
  We test the Sinc function representation, a novel method for
  numerically evaluating Feynman diagrams, by using it to evaluate the
  three-loop master diagrams. Analytical results have been obtained
  for all these diagrams, and we find excellent agreement between our
  calculations and the exact values.  The Sinc function representation
  converges rapidly, and it is straightforward to obtain accuracies of
  1 part in $10^6$ for these diagrams and with longer runs we found
  results better than 1 part in $10^{12}$.  Finally, this paper
  extends the Sinc function representation to diagrams containing
  massless propagators.

\end{abstract}

BROWN-HET-1192

\bigskip
\mbox{}
]


\section{Introduction} 

The Sinc function representation, introduced by us in
Ref.~\cite{EastherET1999c}, is a new approach to numerically evaluating
Feynman diagrams.  This paper tests the Sinc function representation
by using it to calculate the three-loop master diagrams, whose
analytical forms were obtained by Broadhurst in a mathematical {\it
  tour de force\/} \cite{Broadhurst1998a}. We have selected these
diagrams because, in order to make a meaningful test of the Sinc
function representation, we must apply it to diagrams which contain
multiple loops and which have been evaluated analytically, so that we
can compare our results to known exact values. The criteria
``topologically complex'' and ``evaluated analytically'' are not
mutually compatible, but the three-loop master diagrams are
sufficiently complex to allow a stringent test of our methods. In
addition, the master diagrams contain massless lines, so we must start
by extending the Sinc function representation to theories with
massless propagators.

The Sinc function representation performs an {\em ab initio\/}
computation of the corresponding integral, so we make no use of the
special analytic properties of the master diagrams. Consequently, our
success with these diagrams supports our more general claim that the
Sinc function representation is a powerful tool for accurately
evaluating arbitrary complex diagrams.

We believe that our results, which are contain up to 12 or 13
significant figures, represent an unprecedented level of precision for
the numerical evaluation of a non-trivial Feynman integral.
Conventional wisdom holds that multi-dimensional Feynman integrals
must be numerically evaluated with Monte Carlo methods, of which VEGAS
is the canonical implementation \cite{Lepage1978a}.  Monte Carlo
methods usually yield the first few significant figures without too
much difficulty but their efficiency drops extremely rapidly as the
desired accuracy is increased, making highly accurate calculations
effectively impossible.

Obviously, in any realistic situation these diagrams correspond to
higher order corrections, and we are capable of a level of precision
far beyond than that of any experimental data. However, if only the
first few significant figures of the exact result are need the Sinc
function representation can compute these very quickly.  Furthermore,
Broadhurst's analytical treatment of the master diagrams relies on the
PSLQ algorithm \cite{FergusonET1999a} for identifying integer
relationships. Part of input required by this algorithm requires is
accurate numerical evaluation of the diagrams.  By first reducing the
master diagrams to one- (or two-) dimensional integrals, Broadhurst
was able to extract the numerical results he needed.  However, if one
wished to apply the same techniques to diagrams that cannot be reduced
to numerical integrals which are tractable using standard methods, the
Sinc function representation provides a promising alternative approach
for their numerical evaluation.

In the following section we summarize the Sinc function
representation, and show how it can be applied to diagrams with both
massive and massless propagators. In Section 3 we define the
three-loop master diagrams and derive their Sinc function
representations. We evaluate these diagrams in Section 4 and compare
our results with the analytic forms, while in Section 5 we discuss the
implications of our work.

\section{Sinc Function Representation}

The Sinc function representation \cite{EastherET1999c} hinges upon a
novel representation of the propagator and, as the name suggests, is
derived using the properties of the generalized Sinc function,
\be
S_k(h,x) = \frac{\sin{\left[\pi (x-kh)/h\right]}}{\pi(x-kh)/h},
\ee
where the capital $S$ distinguishes this from the more familiar form,
$\mbox{sinc}(x) = \sin(x)/x$.  The properties of the $S_k(h,x)$ are
thoroughly discussed by Stenger \cite{StengerBK1}. For our purposes,
their most important attributes are that any function $f(z)$ which is
analytic on an infinite strip of the complex plane that includes the
real axis can be approximated as
\be
f(z) \approx \sum_{k=-\infty}^{\infty} f(kh) S_k(h,z),
\label{sincexpansion}
\ee
while
\be
\int_{-\infty}^{\infty} {f(z)\, dz } \approx \sum_{k=-\infty}^{\infty}
f(kh).
\label{sincintegral}
\ee
Crucially, the accuracy of the approximation improves exponentially as
$h\rightarrow 0$ \cite{StengerBK1}. In Ref.~\cite{EastherET1999c} we
constructed the Sinc function representation for the scalar field
propagator which, in Euclidean momentum space, has the form
\be
\GLam(p) = \frac{1}{p^2 + m^2} e^{-p^2 /\Lambda^2} ,\label{GLamp} 
\ee
where we have introduced a cutoff in anticipation of divergent
integrals.  We now derive a version of the Sinc function
representation which is applicable to both massive and massless
fields. 

The spacetime propagator is the Fourier transform of $\GLam(p)$,
\bea
\GLam(x) &=& \frac{1}{(4\pi)^2} \int_0^\infty ds \frac{1}{(s^2 +
\frac{1}{\Lambda^2})^2} \times\nonumber \\
 && \qquad \exp{\left(-m^2 s -
\frac{x^2}{s^2 +\frac{1}{\Lambda^2}}\right)} \label{GLam},
\eea
where we introduced $s$ by exponentiating the denominator, and
then performed the Fourier transform. We now replace $s$ by $e^z$ and
use \eq{sincintegral} to approximate the integral over $z$, which
yields the Sinc function representation of the propagator,
\bea
\GLamh(x) &=& \frac{h}{(4\pi)^2 }\sum_{k=-\infty}^{\infty} 
 \frac{\exp{\left(k h - m^2 e^{kh}\right)}}
    {\left(e^{kh} + \frac{1}{\Lambda^2}\right)^2} \times \nonumber \\
   && \qquad \exp{\left(-\frac{x^2}{ e^{kh} + \frac{1}{\Lambda^2}}\right)}. 
\label{GLamh2}
\eea
The massless limit of \eq{GLamh2} is obvious, and setting $m=0$ makes
no difference to its impressive convergence properties
\cite{EastherET1999c}. The difference between this form of $\GLamh$
and the form found in Ref.~\cite{EastherET1999c} is largely aesthetic:
when $m\ne0$, rescaling $s$ to $s/m$ makes all the terms inside the
exponentials explicitly dimensionless.  Lastly, it will be convenient
to write $\GLamh$ as
\be
\GLamh(a-b) =  \frac{h}{(4\pi)^2 }\sum_{k=-\infty}^{\infty}{ 
  p(k) \exp{\left[- \frac{ (a-b)^2}{4 c(k)} \right]}}
\label{GLamh3}
\ee
where 
\begin{eqnarray}
c(k) &=& e^{kh} + \frac{1}{\Lambda^2}, \label{ckdef} \\
p(k) &=& \frac{\exp{\left(kh -m^2 e^{kh}\right)}}
    {(e^{kh} + \frac{1}{\Lambda^2})^2}
         = \frac{\exp{\left(kh-m^2 e^{kh}\right)}}{c(k)^2}. \label{pkdef}
\end{eqnarray}

Feynman diagrams are integrals over products of propagators, and we
derive the Sinc function representation for a given diagram by
inserting $\GLamh$ for each line. All the spacetime integrals required
by the internal vertices can be performed analytically, and we are
left with an $N$-dimensional infinite sum which approximates the
original integral, where $N$ is the number of internal lines in the
diagram.  In Ref.~\cite{EastherET1999c} we give {\em Sinc function Feynman
rules\/} which describe the construction the Sinc function
representation for an arbitrary diagram, and these are not changed by
the inclusion of the massless propagators.

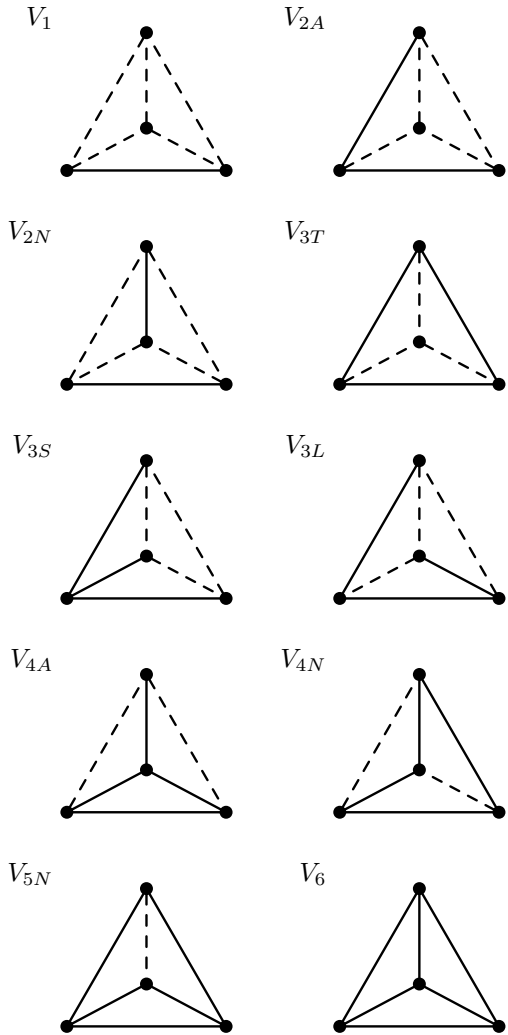
\begin{figure}[tb]
\begin{center}
\begin{fmffile}{tetra}
\begin{fmfgraph*}(100,80)
  \fmfpen{thin}
  \fmfforce{20,10}{v1}
  \fmfforce{80,10}{v2}
  \fmfforce{50,61.96}{v3}
  \fmfforce{50,25.98}{v4}
  \fmfforce{20,60}{v5}
  \fmf{vanilla}{v1,v2}
  \fmf{dashes}{v4,v2,v3,v1,v4,v3}
  \fmfdotn{v}{4}
  \fmflabel{$V_1$}{v5}
\end{fmfgraph*} 
\begin{fmfgraph*}(100,80)
  \fmfpen{thin}
  \fmfforce{20,10}{v1}
  \fmfforce{80,10}{v2}
  \fmfforce{50,61.96}{v3}
  \fmfforce{50,25.98}{v4}
  \fmfforce{20,60}{v5}
  \fmf{vanilla}{v2,v1,v3}
  \fmf{dashes}{v1,v4,v3,v2,v4}
  \fmfdotn{v}{4}
  \fmflabel{$V_{2A}$}{v5}
\end{fmfgraph*} 
\begin{fmfgraph*}(100,80)
  \fmfpen{thin}
  \fmfforce{20,10}{v1}
  \fmfforce{80,10}{v2}
  \fmfforce{50,61.96}{v3}
  \fmfforce{50,25.98}{v4}
  \fmfforce{20,60}{v5}
  \fmf{vanilla}{v1,v2}
  \fmf{dashes}{v4,v1,v3,v2,v4}
  \fmf{vanilla}{v3,v4}
  \fmfdotn{v}{4}
  \fmflabel{$V_{2N}$}{v5}
\end{fmfgraph*}
\begin{fmfgraph*}(100,80)
  \fmfpen{thin}
  \fmfforce{20,10}{v1}
  \fmfforce{80,10}{v2}
  \fmfforce{50,61.96}{v3}
  \fmfforce{50,25.98}{v4}
  \fmfforce{20,60}{v5}
  \fmf{vanilla}{v1,v2,v3,v1}
  \fmf{dashes}{v1,v4,v2}
  \fmf{dashes}{v3,v4}
  \fmfdotn{v}{4}
  \fmflabel{$V_{3T}$}{v5}
\end{fmfgraph*}
\begin{fmfgraph*}(100,80)
  \fmfpen{thin}
  \fmfforce{20,10}{v1}
  \fmfforce{80,10}{v2}
  \fmfforce{50,61.96}{v3}
  \fmfforce{50,25.98}{v4}
  \fmfforce{20,60}{v5}
  \fmf{vanilla}{v1,v2}
  \fmf{vanilla}{v4,v1,v3}
  \fmf{dashes}{v2,v3,v4,v2}
  \fmfdotn{v}{4}
  \fmflabel{$V_{3S}$}{v5}
\end{fmfgraph*}
\begin{fmfgraph*}(100,80)
  \fmfpen{thin}
  \fmfforce{20,10}{v1}
  \fmfforce{80,10}{v2}
  \fmfforce{50,61.96}{v3}
  \fmfforce{50,25.98}{v4}
  \fmfforce{20,60}{v5}
  \fmf{vanilla}{v3,v1,v2,v4}
  \fmf{dashes}{v2,v3,v4,v1}
  \fmfdotn{v}{4}
  \fmflabel{$V_{3L}$}{v5}
\end{fmfgraph*}
\begin{fmfgraph*}(100,80)
  \fmfpen{thin}
  \fmfforce{20,10}{v1}
  \fmfforce{80,10}{v2}
  \fmfforce{50,61.96}{v3}
  \fmfforce{50,25.98}{v4}
  \fmfforce{20,60}{v5}
  \fmf{vanilla}{v3,v4,v1,v2,v4}
  \fmf{dashes}{v1,v3,v2}
  \fmfdotn{v}{4}
  \fmflabel{$V_{4A}$}{v5}
\end{fmfgraph*}
\begin{fmfgraph*}(100,80)
  \fmfpen{thin}
  \fmfforce{20,10}{v1}
  \fmfforce{80,10}{v2}
  \fmfforce{50,61.96}{v3}
  \fmfforce{50,25.98}{v4}
  \fmfforce{20,60}{v5}
  \fmf{vanilla}{v2,v1,v4,v3,v2}
  \fmf{dashes}{v1,v3}
  \fmf{dashes}{v2,v4}
  \fmfdotn{v}{4}
  \fmflabel{$V_{4N}$}{v5}
\end{fmfgraph*}
\begin{fmfgraph*}(100,80)
  \fmfpen{thin}
  \fmfforce{20,10}{v1}
  \fmfforce{80,10}{v2}
  \fmfforce{50,61.96}{v3}
  \fmfforce{50,25.98}{v4}
  \fmfforce{20,60}{v5}
  \fmf{vanilla}{v1,v2,v3,v1,v4,v2}
  \fmf{dashes}{v3,v4}
  \fmfdotn{v}{4}
  \fmflabel{$V_{5N}$}{v5}
\end{fmfgraph*}
\begin{fmfgraph*}(100,80)
  \fmfpen{thin}
  \fmfforce{20,10}{v1}
  \fmfforce{80,10}{v2}
  \fmfforce{50,61.96}{v3}
  \fmfforce{50,25.98}{v4}
  \fmfforce{20,60}{v5}
  \fmf{vanilla}{v1,v2,v4,v1,v3,v4}
  \fmf{vanilla}{v2,v3}
  \fmfdotn{v}{4}
  \fmflabel{$V_{6}$}{v5}
\end{fmfgraph*}
\end{fmffile}
\end{center}
\caption[]{The three-loop master diagrams, where dashed lines are
massless. \label{diag}}
\end{figure}

\begin{figure}[tb]
\begin{center}
\begin{fmffile}{label}
\begin{fmfgraph*}(100,80)
  \fmfpen{thin}
  \fmfforce{20,10}{v1}
  \fmfv{label=$x_1$}{v1}
  \fmfforce{80,10}{v2}
  \fmfv{label=$x_2$}{v2}
  \fmfforce{50,61.96}{v3}
  \fmfv{label=$x_3$}{v3}
  \fmfforce{50,25.98}{v4}
  \fmfv{label=$x_4$}{v4}
  \fmfforce{20,60}{v5}
  \fmf{vanilla}{v1,v2}
  \fmf{vanilla}{v4,v2,v3,v1,v4,v3}
  \fmfdotn{v}{4}
\end{fmfgraph*} 
\end{fmffile}
\end{center}
\caption{The generic topology of the master diagram, showing our
choice of vertex labels. \label{vert}}  
\end{figure}
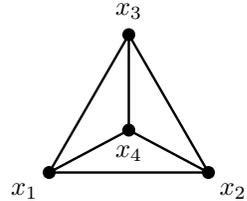

\section{The Master Diagrams}

The 10 three-loop master diagrams, or vacuum tadpoles, are depicted in
\fig{diag}. In addition to their intrinsic beauty, their practical use
is that three-loop diagrams dominated by a single, large mass and with
small external momenta can be expressed in terms of the master
diagrams.  Four of these diagrams were given analytically before 1998
\cite{ChetyrkinET1981a,Tkachov1981a,Broadhurst1992a,Avdeev1996a}, and
Broadhurst found results for the remaining six \cite{Broadhurst1998a}.
Applying the coordinate space Feynman rules to the overall
``tetrahedron'' topology of the master diagrams yield the following
integral
\begin{eqnarray}
V_j(\Lambda) &=& (4\pi)^6 \int d^4 x_1 d^4 x_2 d^4 x_3 G_1(x_1-x_2)
\nonumber \\ && \quad G_2(x_1-x_3) G_3(x_1-x_4) G_4(x_2-x_3) \nonumber
\\ && \qquad G_5(x_2-x_4) G_6(x_3-x_4). \label{vj}
\end{eqnarray}
The subscript $j$ labels the specific combination of massive and
massless propagators, as enumerated in \fig{diag}, while the prefactor
reconciles our measure with Broadhurst's. \Eq{vj} is a coordinate
space integral, but this analysis works equally well in momentum
space.  The vertex labels are shown in \fig{vert}, and from here on we
use Lorentz invariance to set $x_4$ to zero. The six propagators are
functions of the mass, $m_j$, carried by each line.  For the master
diagrams, the $m_j$ are either $0$ or $1$, but the Sinc function
representation yields results even if all the non-zero masses are
different. The integral $V_j$ is divergent, but it is regulated by
employing the cut-off propagator, $\GLam(x)$.

Applying the Sinc function Feynman rules, we insert $\GLamh$ for each
of the propagators, and perform the Gaussian integrals. After some
algebraic manipulation, we obtain the the Sinc function representation 
of the general integral $V_j$,
\be
V_j(\Lambda,h)= h^6 \sum_{k} 
 \frac{\prod_{n=1}^{6} c_n(k_n)^2 p_n(k_n)}{(a + b c_6(k_6))^2},
\label{sincrep}
\ee
where
\bea
a &=& (c_2(k_2)+ c_5(k_5))c_3(k_3)c_4(k_4) + \nonumber \\
&& c_2(k_2) c_5(k_5) (c_3(k_3) + c_4(k_4)) + \nonumber \\
&& c_1(k_1) (c_2(k_2)+c_3(k_3)) (c_4(k_4), + c_5(k_5)) \\
b &=& (c_2(k_2) + c_4(k_4))(c_3(k_3)+c_5(k_5)) + \nonumber \\
 && c_1(k_1)(c_2(k_2) + c_3(k_3) + c_4(k_4) +c_5(k_5)).
\eea
The $c_i$ and $p_i$ are defined by \eqs{ckdef}{pkdef} and the
summation over $k$ is shorthand for infinite sums running between
$\pm\infty$ for each $k_i$, $i=1\cdots6$.

To compute $V_j$ from the Sinc function representation we numerically
evaluate the six-dimensional infinite sum in \eq{sincrep}. In
practice, $V_j(\Lambda,h)$ is an approximation to $V_j(\Lambda)$,
which is exact in the limit $h\rightarrow0$ \cite{EastherET1999c}. By
choosing $h$, we effectively fix the accuracy with which it is possible
to evaluate $V_j$. Moreover, although the cutoff renders $V_j$ finite,
the result of evaluating \eq{sincrep} directly depends on $\Lambda$,
and we must remove this regularization dependence before proceeding.

Examining the six-dimensional sum, we see that if $\Lambda
\rightarrow 0$ and $k_i = k$, then the general term approaches a
constant, $h^2/(256 (4\pi)^6)$, as $k$ becomes large and
negative. Consequently, the sum diverges in the limit $\Lambda
\rightarrow \infty$, and we have thus identified the origin of the
divergent part of $V_j$ in the Sinc function representation.

In Ref.~\cite{EastherET1999c}, we renormalized the Sinc function
representation of the sunset graph by expanding it about its external
momentum and subtracting the $p^0$ and $p^2$ terms.  This approach
cannot be used with the master diagrams as they do not carry external
momenta. However, the divergent part does not depend on any of the
masses in the diagram. Consequently, the difference of any two
$V_j(\Lambda,h)$ is finite even when $\Lambda\rightarrow\infty$, as
the divergent parts in the sums for each of the $V_j(\Lambda,h)$ will
cancel term by term. In an approach that parallels Broadhurst's
analytical treatment of the $V_j$, we write the differences between
$V_j$ and $V_0$ as
\be
F_j(h) = \sum_{k_i} \lim_{\Lambda\rightarrow\infty}(t_1 - t_j)
\label{Fj}
\ee where $t_j$ is the general term from the six-dimensional sum that
gives the Sinc function representation of $V_j$.  By forming the
difference between $V_1$ and $V_j$ in this way, and {\em then\/}
taking the limit $\Lambda\rightarrow\infty$, we compute the finite
$F_j$ without ever needing to evaluate the $V_j(\Lambda)$, which
diverge in the limit $\Lambda\rightarrow\infty$. By taking this
approach we avoid the loss of precision that would result from
attempting to evaluate $F_j$ by computing $V_j$ at large but finite
$\Lambda$ and then subtracting the result from $V_1$.

In \fig{diag} we have used the symmetry of the diagrams to ensure that
$G_1$, the propagator between $x_1$ and $x_2$, is always massive.
Since all the other lines in $V_1$ are massless and $p_k$ with $m^2>0$
is always less than $p_k$ with $m^2=0$, our choice of labeling ensures
that $t_1>t_j$ for any $\{k_1,\cdots,k_6\}$.  Thus all the terms in
$F_j$ are positive, and a numerical evaluation of the multidimensional
sum approaches the limiting value monotonically.

We could be content with simply calculating the differences between
diagrams and comparing our result to the analytic values. However, we
can make use of a single analytical result, namely that the finite
part of $V_1$ is $6 \zeta(3) + 3 \zeta(4)$, to relate our Sinc
function representation calculation of the $F_j$ with the renormalized
$V_j$.

\begin{table}[tb]
\begin{tabular}{|r||l|l|l|}
\hline 
$j$ & $F_j$ (exact) &  $F_j$ (computed)  & Error \\
\hline
\hline  
$2A$ &  8.65858586969 &  8.65857982635  & $-6.98\,{{10}^{-7}}$  \\
$2N$ &  9.33906272305 &   9.33905792236   & $ -5.14\,{{10}^{-7}}$ \\
$3T$ &  12.9878788045 &  12.98787117   &  $-5.88\,{{10}^{-7}}$  \\
$3S$ &  13.3201733442 &  13.3201656342  & $-5.79\,{{10}^{-7}}$ \\
$3L$ &   13.4863206141 &  13.4863128662   & $ -5.74\,{{10}^{-7}}$ \\
$4A$ &  16.372515904 &    16.3725070953  &  $ -5.38\,{{10}^{-7}}$\\
$4N$ &  16.5134789787 & 16.513469696  &  $-5.62\,{{10}^{-7}}$ \\
$5N$ &  18.6761709376 &    18.6761627197 &$     -4.4\,{{10}^{-7}}$  \\
$6$ &  20.4945895999  &  20.4945793152 & $-5.02\,{{10}^{-7}}$\\
\hline
\end{tabular}
\caption{We have computed the 9 non-trivial $F_j$ and compared them to 
the exact values.  These results were obtained with $h=0.7$. \label{table}}
\end{table}

\begin{figure}[tbp]
\begin{center}
\begin{tabular}{c}
\epsfysize=4cm
\epsfbox{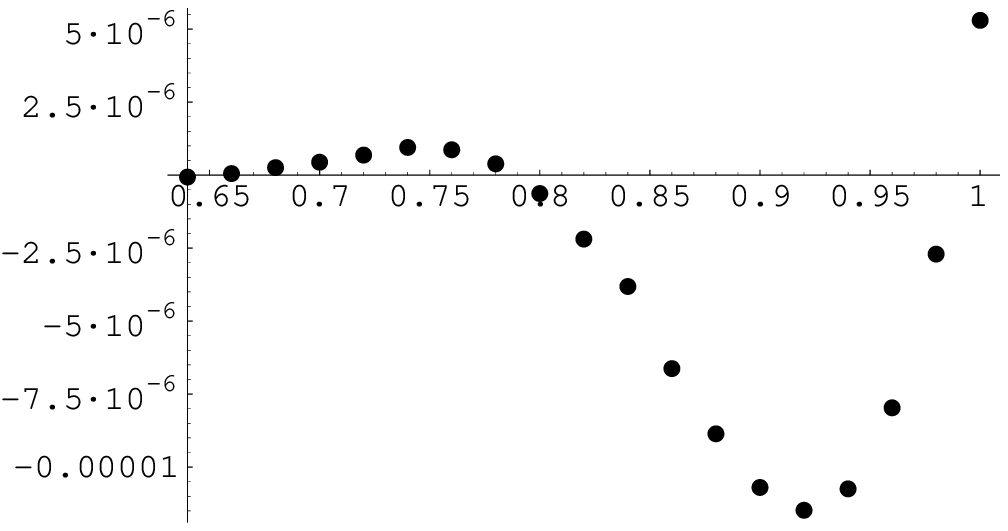} \\
\epsfysize=4cm
\epsfbox{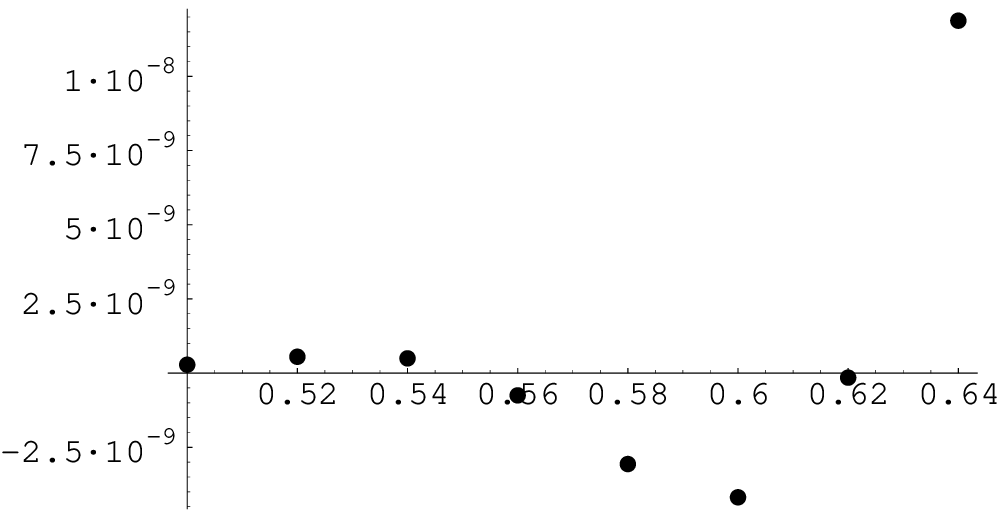} 
 \end{tabular}
\end{center}
\caption[]{We plot $1-F_{10}(h)/F_{10}$, as a function of $h$. \label{hdepend}}
\end{figure}

\section{Numerical Evaluations}

We evaluate the $F_j$ with a code that is similar to those used to
obtain the results given in Ref.~\cite{EastherET1999c}.\footnote{Sample
codes for calculating $F_j$ can be found at the URL \\ {\tt
http://www.het.brown.edu/people/easther/feynman/}}
The code adaptively limits itself to the region of the six-dimensional
lattice defined by the $k_i$ that make the strongest contribution to
the overall sum. We accelerate the convergence of each of the six sums
by using an Aitken $\delta^2$ extrapolation
\cite{EastherET1999c,PressBK1}. This can significantly reduce the time
needed to evaluate the sums, but the problem is still tractable
without the extrapolation.  While we have endeavored to ensure that
the algorithm is implemented efficiently, there is no guarantee that
it is the optimal method for evaluating the Sinc function
representation of $F_j$.  Consequently, the timing data we present
(obtained using a 250MHz Ultra{\sc sparc } II CPU) should be
interpreted as a upper limit to the effort needed to evaluate these
diagrams' Sinc function representations.

When using \eq{Fj} to evaluate the $F_j$, we find that there are two
distinct sources of error. The first is the intrinsic difference
between $F_j(h)$ and the exact value of $F_j$, and the second is the
numerical error (if any) in the evaluation of the infinite series.
This has two possible origins: either because the summation was
truncated prematurely, or because of an accumulation of numerical
error.  In Table~\ref{table}, we show the results for the nine
non-trivial $F_j$, with $h=0.7$. It is comparatively easy for us to
obtain results accurate to within a few parts in $10^7$.

The difference between the Sinc function representation and the exact
value decreases quickly as $h$ is reduced.  In \fig{hdepend}, we plot
the error in the value of $F_{10}$ computed from the Sinc function
representation, as a function of $h$. The effort required to evaluate
the sum increases at smaller $h$. First, the general term dies away
more slowly when $h$ is decreased, so we must add more terms to
evaluate the Sinc function representation to a given level of
accuracy. In addition, with smaller values of $h$, the maximum size of
the terms which can be ignored by the numerical algorithm also drops,
which further increases the effort needed to compute the
multi-dimensional sums.  With $h\sim1.0$, the Sinc function
representation could be evaluated in approximately three minutes on a
workstation. The exact timings depend on the truncation procedure used
in our code, which is somewhat heuristic and probably too
conservative.  With $h\sim 0.7$, the evaluation time rises to around 20
minutes, and with $h\sim0.5$, several hours are required.

By paying close attention to the form of the sum, it is possible to
improve on these speeds. We are particularly interested in computing
$F_6$, the master diagram in which all lines are massive. A direct
analytical treatment of this diagram yields a double integral, which
must then be computed numerically in order to verify the results
obtained with the PSLQ algorithm \cite{Broadhurst1998a}. All the other
results identified this way can be compared to (at worst) one
dimensional integrals, which could be computed to very high precision.
However, the result for $F_6$ has not been checked beyond 15 decimal
places.

We can make two improvements to the efficiency of the Sinc function
representation of $F_6$. First, we can compute $V_{5N} - V_{6}$,
instead of $F_6$, as the sums in the positive $k_i$ directions for
$i=2,\cdots,5$ converge much more rapidly than those in $F_6$, due to
the $\exp{(-e^{k_i h})}$ terms contributed by the extra massive
propagators in $V_{5N}$.  Since $V_{5N}$ is known analytically, we can
then deduce both $V_6$ and $F_6$. Second, we can speed the
evaluation of the inner loop since we can write \eq{sincrep} as
\be
V_j(\Lambda,h) =  h^6 \sum_{k_1\cdots k_5}\left[ 
 \frac{\prod_{n=1}^{5} c_n(k_n)^2 p_n(k_n)}{b^2}  f(\frac{a}{b})\right]
\ee
where
\be
f(\frac{a}{b}) = \sum_{k_6}\frac{c_6(k_6)^2 p_6(k_6)}{(a/b +  c_6(k_6))^2},
\ee
and we recall that $b$ does not depend on $k_6$. Since $f(a/b)$
depends on the single parameter $a/b$ we can construct an
interpolation table which allows us to evaluate $f$ in far less time
than it takes to compute the sum over $k_6$. This effectively reduces
the problem to a five-dimensional sum. The combination of these two
methods improves the evaluation time by an order of magnitude or more,
and it is straightforward to calculate $V_6$ to within a few parts in
$10^{13}$.

Unfortunately, going beyond 16 digits of precision requires the use of
quadruple precision variables, and this dramatically lowers the
computational efficiency of the code on present hardware.  However,
given a machine that performed 128-bit numerical arithmetic at similar
speeds to 64-bit arithmetic, we would be able to evaluate $F_6$ to a
precision much better than 1 part in $10^{-16}$.

\section{Discussion}

The immediate purpose of this paper was to show that the Sinc function
representation reproduces the analytic results for the three-loop
master diagrams. Since our results are limited only by the finite
accuracy of double-precision computer arithmetic, we have emphatically
demonstrated the efficacy of the Sinc function representation in this
specific case.  We pursued this problem to test the usefulness of the
Sinc function representation as a tool for evaluating more general
Feynman integrals. The Sinc function representation does not rely on
any special properties of the diagrams being evaluated, and we made no
use of the analytic knowledge gained from previous work on the master
diagrams.  Thus, these results support our claim
that the Sinc function representation may be a useful approach to
evaluating general higher order diagrams \cite{EastherET1999c}.

There is considerable theoretical interest in performing accurate
calculations of specific diagrams to assist the investigation of
non-trivial relationships between diagrams.  Broadhurst's evaluation
of the three-loop master diagrams is a prime example of this type of
work, which has the ability to illuminate the deep structure of
perturbative quantum field theory.  The Sinc function representation
has the potential to facilitate this approach, since it is fast,
accurate, and does not require a partial analytical evaluation of the
diagram.

When analyzing experimental data, we are unlikely to need the high
levels of accuracy achieved in this paper.  However, if we only want a
few significant figures, we can increase $h$, which allows us to
evaluate diagrams very rapidly.  The Sinc function representation's
convergence properties, lack of analytic overhead, and applicability
to arbitrary topologies suggests that it may be a useful tool for
automatically calculating large sets of diagrams.  However, before
this possibility can be explored in detail, we must generalize the Sinc
function representation to fermionic and vector fields, and describe a
renormalization procedure that can be applied to the Sinc function
representation of an arbitrary diagram. This work is currently in
progress.

Finally, while Feynman integrals arise in perturbative quantum field
theory, the Sinc function representation is an outgrowth of a new
approach to ``exact'' numerical quantum field theory, the {\em Source
Galerkin\/} method
\cite{GarciaET1994a,%
GarciaET1996a,%
LawsonET1996a,%
Hahn1998a,%
HahnET1999a,%
EmirdagET1999a}.
To apply the source Galerkin method we must carry out integrals over
products of terms proportional to the free propagator, which resemble
Feynman integrals. Consequently, the numerical codes and analytical
insight needed to apply the Sinc function representation to Feynman
integrals also improve the efficiency of the Source Galerkin method.

\section*{Acknowledgments}

We thank Gyan Bhanot for a useful discussion.  Computational work in
support of this research was performed at the Theoretical Physics
Computing Facility at Brown University.  This work is supported by DOE
contract DE--FG0291ER40688, Tasks A and D.


\begin{thebibliography}{10}

\bibitem{EastherET1999c}
R. Easther, G. Guralnik, and S. Hahn, \\hep-ph/9903255  (1999).

\bibitem{Broadhurst1998a}
D. Broadhurst, Eur. Phys. J. {\bf C8},  311  (1999).

\bibitem{Lepage1978a}
G.~P. Lepage, J. Comput. Phys. {\bf 27},  192  (1978).

\bibitem{FergusonET1999a}
H.~R.~P. Ferguson, D.~H. Bailey, and S. Arno, Math. Comput. {\bf 68},  351
  (1999).

\bibitem{StengerBK1}
F. Stenger, {\em Numerical Methods Based on Sinc and Analytic Functions}
  (Springer Verlag, New York, NY, 1993).

\bibitem{ChetyrkinET1981a}
K.~G. Chetyrkin, A.~L. Kataev, and F.~V. Tkachov, Phys. Lett. B {\bf 99},  147
  (1981).

\bibitem{Tkachov1981a}
F.~V. Tkachov, Phys. Lett. B {\bf 100},  65  (1981).

\bibitem{Broadhurst1992a}
D. Broadhurst, Z. Phys. C {\bf 54},  599  (1992).

\bibitem{Avdeev1996a}
L.~V. Avdeev, Comput. Phys. Commun. {\bf 98},  15  (1996).

\bibitem{PressBK1}
W. Press, S. Teukolsky, W. Vetterling, and B. Flannery, {\em Numerical Recipes
  in Fortran}, 2 ed. (Cambridge UP, Cambridge, 1992).

\bibitem{GarciaET1994a}
S. Garc\'{\i}a, G.~S. Guralnik, and J.~W. Lawson, Phys. Lett. B {\bf 333},  119
   (1994).

\bibitem{GarciaET1996a}
S. Garc\'{\i}a, Z. Guralnik, and G.~S. Guralnik, hep-th/9612079  (1996).

\bibitem{LawsonET1996a}
J.~W. Lawson and G.~S. Guralnik, Nucl. Phys. B {\bf 459},  589  (1996).

\bibitem{Hahn1998a}
S.~C. Hahn, Ph.D. thesis, Brown University, 1998.

\bibitem{HahnET1999a}
S.~C. Hahn and G.~S. Guralnik, hep-th/9901019  (1999).

\bibitem{EmirdagET1999a}
P. Emirda\u{g}, R. Easther, G. Guralnik, and S. Hahn, hep-lat/9909122  (1999).

\end{thebibliography}

\end{document}